%% file: top.tex
\documentclass[sigconf,authorversion]{acmart}

\usepackage{microtype}      
\usepackage{enumitem}

\usepackage[ruled]{algorithm2e} 
\usepackage{subcaption}

\SetAlFnt{\small}
\SetAlCapFnt{\small}
\SetAlCapNameFnt{\small}
\SetAlCapHSkip{0pt}

\acmJournal{TOG}

\newcommand{\rev}[1]{{#1}}

\copyrightyear{2023}
\acmYear{2023}
\setcopyright{acmlicensed}\acmConference[SA Conference Papers '23]{SIGGRAPH Asia 2023 Conference Papers}{December 12--15, 2023}{Sydney, NSW, Australia}
\acmBooktitle{SIGGRAPH Asia 2023 Conference Papers (SA Conference Papers '23), December 12--15, 2023, Sydney, NSW, Australia}
\acmPrice{15.00}
\acmDOI{10.1145/3610548.3618233}
\acmISBN{979-8-4007-0315-7/23/12}

\begin{document}
\title{Diffusion-based Holistic Texture Rectification and Synthesis}

\author{Guoqing Hao}
\orcid{0000-0002-6132-8830}
\affiliation{%
 \institution{University of Tsukuba}
 \city{Tsukuba}
 \state{Ibaraki}
 \postcode{3050085}
 \country{Japan}
}
\affiliation{%
 \institution{National Institute of Advanced Industrial Science and Technology (AIST)}
 \city{Tsukuba}
 \state{Ibaraki}
 \postcode{3058560}
 \country{Japan}}
\email{hao\_guoqing@cvlab.cs.tsukuba.ac.jp}

\author{Satoshi Iizuka}
\orcid{0000-0001-9136-8297}
\affiliation{%
 \institution{University of Tsukuba}
 \city{Tsukuba}
 \state{Ibaraki}
 \postcode{3050085}
 \country{Japan}
}
\email{iizuka@cs.tsukuba.ac.jp}

\author{Kensho Hara}
\orcid{0000-0001-6463-7738}
\affiliation{%
\institution{National Institute of Advanced Industrial Science and Technology (AIST)}
\city{Tsukuba}
\state{Ibaraki}
\postcode{3058560}
\country{Japan}}
\email{kensho.hara@aist.go.jp}

\author{Edgar Simo-Serra}
\orcid{0000-0003-2544-8592}
\affiliation{%
 \institution{Waseda University}
 \city{Shinjuku}
 \state{Tokyo}
 \postcode{1698050}
 \country{Japan}
}
\email{ess@waseda.jp}

\author{Hirokatsu Kataoka}
\orcid{0000-0001-8844-165X}
\affiliation{%
 \institution{National Institute of Advanced Industrial Science and Technology (AIST)}
 \city{Tsukuba}
 \state{Ibaraki}
 \postcode{3058560}
 \country{Japan}}
\email{hirokatsu.kataoka@aist.go.jp}

\author{Kazuhiro Fukui}
\orcid{0000-0002-4201-1096}
\affiliation{%
 \institution{University of Tsukuba}
 \city{Tsukuba}
 \state{Ibaraki}
 \postcode{3050085}
 \country{Japan}
}
\email{kfukui@cs.tsukuba.ac.jp}

\renewcommand\shortauthors{Hao, G. et al}

\begin{teaserfigure}
    \includegraphics[width=\linewidth]{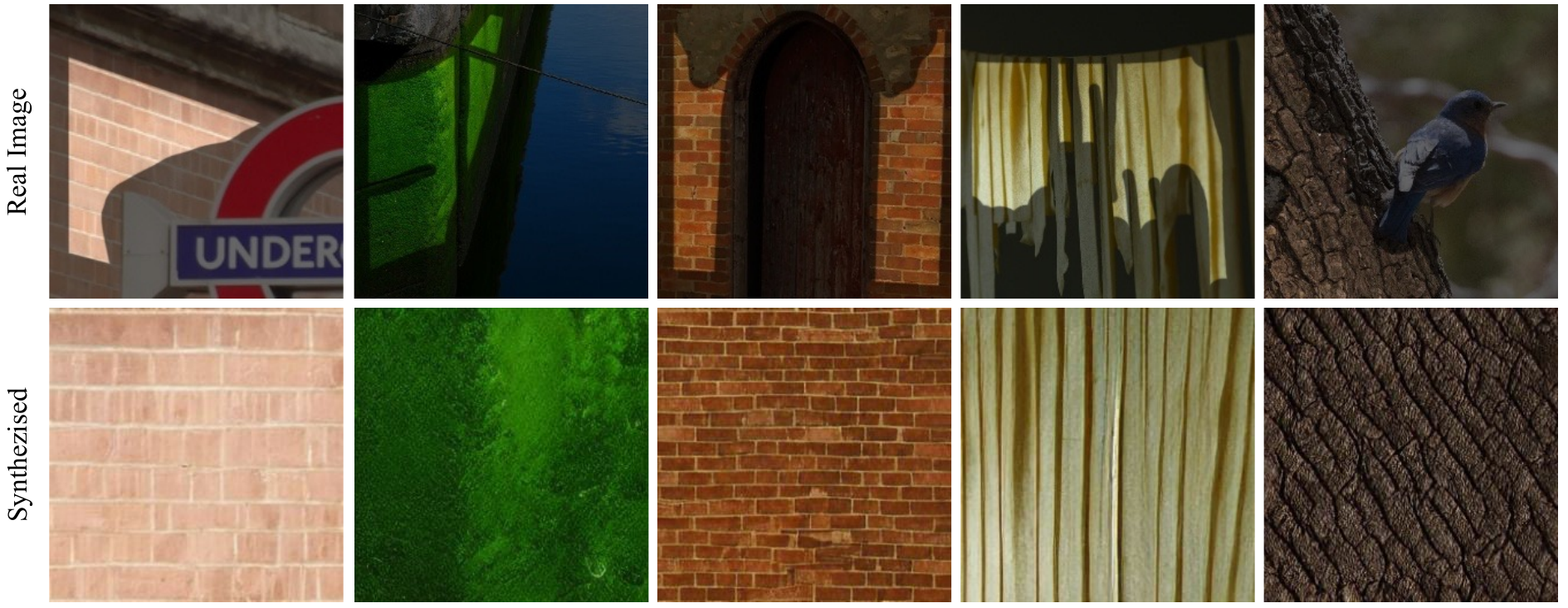}
    \vspace{-7mm}
    \caption{\textbf{Rectified and synthesized texture results from real
    images.} Our framework is able to rectify degraded textures, which include
    occlusions and geometric deformations, and synthesize holistic textures from
    selected areas. The first row presents real images, where the masked input
    to our framework is highlighted, while the second row shows outputs
    generated by our proposed approach. Our method accomplishes more than
    merely filling in missing regions; it also rectifies geometric
    deformations, including perspective variations and distortions to
    synthesize textures amenable for usage in many different applications such
    as 3D modelling. \rev{Photographs courtesy of \href{https://flic.kr/p/xTGFTZ}{Elliott Brown} (CC-BY), \href{https://flic.kr/p/7VZDyG}{Scott Meis} (CC-BY), \href{https://flic.kr/p/MBcQRT}{denisbin} (Public Domain), \href{https://flic.kr/p/9aLQiV}{Jameel Winter} (CC-BY), and \href{https://flic.kr/p/T7GtTA}{Bettina Arrigoni} (CC-BY).}}
    \label{fig:teaser}
\end{teaserfigure}

\begin{abstract}
We present a novel framework for rectifying occlusions and distortions in
degraded texture samples from natural images. Traditional texture synthesis
approaches focus on generating textures from pristine samples, which
necessitate meticulous preparation by humans and are often unattainable in most
natural images. These challenges stem from the frequent occlusions and
distortions of texture samples in natural images due to obstructions and
variations in object surface geometry. To address these issues, we propose a
framework that synthesizes holistic textures from degraded samples in natural images,
extending the applicability of exemplar-based texture synthesis techniques. Our
framework utilizes a conditional Latent Diffusion Model (LDM) with a novel
occlusion-aware latent transformer. This latent transformer not only
effectively encodes texture features from partially-observed samples necessary
for the generation process of the LDM, but also explicitly captures long-range
dependencies in samples with large occlusions. To train our model, we introduce a
method for generating synthetic data by applying geometric transformations and free-form mask generation to
clean textures. Experimental results demonstrate that our framework
significantly outperforms existing methods both quantitatively and
quantitatively. Furthermore, we conduct comprehensive ablation studies to
validate the different components of our proposed framework. Results are
corroborated by a perceptual user study which highlights the efficiency of our
proposed approach.
\end{abstract}

%
%

\ccsdesc[500]{Computing methodologies~Image processing}
\ccsdesc[300]{Computing methodologies~Texturing}

%
%

\keywords{Texture rectification, texture synthesis, diffusion models}

\maketitle

\input{01_intro}

\input{02_related}
\input{03_method}
\input{04_exps}
\input{05_conc}

\begin{acks}
This work was partially supported by JST SPRING (Hao, Grant Number: JPMJSP2124), JST PRESTO (Iizuka, Grant Number: JPMJPR21C1), and JSPS KAKENHI (Hara, Grant Number: JP21H04908).
\end{acks}

\bibliographystyle{ACM-Reference-Format}
\bibliography{top}

\newpage

\newcommand{\rfig}[1]{%
\includegraphics[width=0.18\linewidth]{figs/synthetic_results/input/#1}&
   \includegraphics[width=0.18\linewidth]{figs/synthetic_results/mat/#1}&
   \includegraphics[width=0.18\linewidth]{figs/synthetic_results/vqgan/#1}&
   \includegraphics[width=0.18\linewidth]{figs/synthetic_results/ours/#1}&
   \includegraphics[width=0.18\linewidth]{figs/synthetic_results/gt/#1}}

\begin{figure*}[t]
   \centering
   \setlength{\tabcolsep}{1pt}
   \begin{tabular}{ccccc}
      \rfig{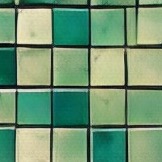} \\
      \rfig{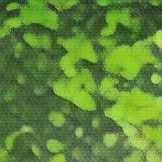} \\
      Input & MAT~\cite{li2022mat} & VQGAN~\cite{esser2021taming} & Ours & Ground Truth \\
   \end{tabular}
   \vspace{-4mm}
   \caption{\textbf{Rectification results on the synthetic test dataset.} Our framework can generate texture images that are perceptually closer to the ground truth than other methods. MAT generates textures that are not related to the input as it falls into the mode-collapse problem. Although the VQGAN can rectify degraded textures, it loses details of the input texture. \rev{Texture images from \cite{dai2014synthesizability}.} }
   \label{fig:synthetic_comp}
\end{figure*}

\newcommand{\ltfig}[1]{%
\includegraphics[width=0.18\linewidth]{figs/ablation_results/input/#1}&
   \includegraphics[width=0.18\linewidth]{figs/ablation_results/pce/#1}&
   \includegraphics[width=0.18\linewidth]{figs/ablation_results/sae/#1}&
   \includegraphics[width=0.18\linewidth]{figs/ablation_results/ours/#1}&
   \includegraphics[width=0.18\linewidth]{figs/ablation_results/gt/#1}}
\newcommand{\condsfig}[1]{%
\includegraphics[width=0.18\linewidth]{figs/ablation_results/input/#1}&
   \includegraphics[width=0.18\linewidth]{figs/ablation_results/concat/#1}&
   \includegraphics[width=0.18\linewidth]{figs/ablation_results/crossattn/#1}&
   \includegraphics[width=0.18\linewidth]{figs/ablation_results/ours/#1}&
   \includegraphics[width=0.18\linewidth]{figs/ablation_results/gt/#1}}

   \begin{figure*}[ht]
   \begin{subfigure}{\linewidth}
         \centering
         \setlength{\tabcolsep}{1pt}
         \begin{tabular}{ccccc}
            \ltfig{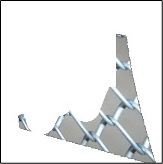} \\
            Input & PCE & SAE & Ours & Ground Truth \\
         \end{tabular}
         \vspace{-3mm}
         \caption{Ablation results on the architecture of the occlusion-aware latent transformer.}
         \label{fig:ab_lt}
   \end{subfigure}
   \vspace{-3mm}
   \begin{subfigure}{\linewidth}
    \setlength{\tabcolsep}{1pt}
    \centering
    \begin{tabular}{ccccc}
      \condsfig{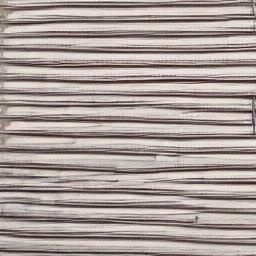} \\
      \condsfig{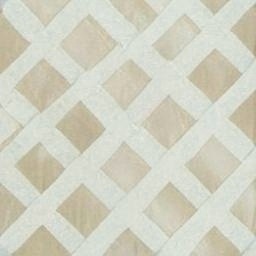} \\
      Input & Concat & Cross-attention & Ours & Ground Truth \\
   \end{tabular}
   \vspace{-3mm}
   \caption{Ablation results on the architecture of the conditioning mechanisms.}
   \label{fig:ab_conds}
   \end{subfigure}
   \vspace{-4mm}
   \caption{\textbf{Visual results of the ablation study.} Fig.~\ref{fig:ab_conds} shows ablation results relating to the architecture of the conditioning mechanisms, while Fig.~\ref{fig:ab_lt} presents ablation results on the architecture of the occlusion-aware latent transformer. It is readily apparent that our full method generates more realistic textures than the other methods. \rev{Texture images from \cite{cimpoi14describing}.}}
\end{figure*}

\newpage

\newcommand{\rtfig}[1]{%
\includegraphics[width=0.24\linewidth]{figs/real_results/input/#1}&
   \includegraphics[width=0.24\linewidth]{figs/real_results/mat/#1}&
   \includegraphics[width=0.24\linewidth]{figs/real_results/vqgan/#1}&
   \includegraphics[width=0.24\linewidth]{figs/real_results/ours/#1}}

\begin{figure*}[t]
   \centering
   \setlength{\tabcolsep}{1pt}
   \begin{tabular}{ccccc}
      \rtfig{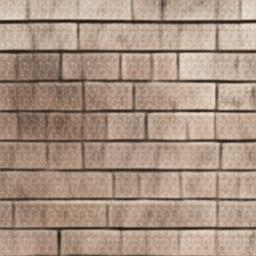} \\
      \rtfig{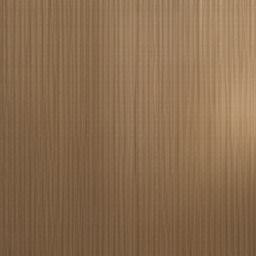} \\
      \rtfig{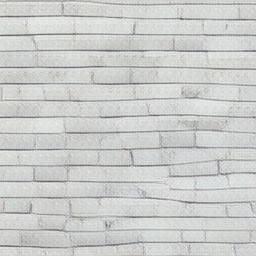} \\
      \rtfig{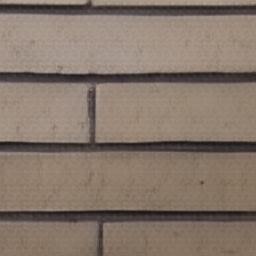} \\
      Real Image & MAT & VQGAN & Ours \\
   \end{tabular}
   \setlength{\tabcolsep}{6pt}
   \vspace{-3mm}
   \caption{\textbf{Rectified and synthesized texture results on real images.} Our framework can generate texture images from real images. One can easily get a planar texture image by selecting desired regions with brush tools. The input to our framework is highlighted. Compared to other methods, our framework preserves the original appearance and synthesizes holistic texture images. \rev{Photographs courtesy of \href{https://flic.kr/p/2juzW5N}{TheTurducken} (CC-BY), \href{https://flic.kr/p/6Wkez7}{Alan Light} (CC-BY), \href{https://flic.kr/p/z6ko7s}{Andrea Dufrenne} (CC-BY), and \href{https://flic.kr/p/YtWrj}{Toshiyuki IMAI} (CC-BY).}}
   \label{fig:real_comp}
\end{figure*}

\end{document}

%% file: 01_intro.tex
\section{Introduction}
Textures are a crucial visual aspect of real-world scenes, representing surface
appearance and consisting of repeating patterns with some inherent randomness.
There are numerous applications in computer graphics and vision that use
textures, including 3D modelling, image editing~\cite{criminisi2003object},
virtual object creation~\cite{chen2018virtual}, and augmented
reality~\cite{isoyama2021effects}. Textures can be derived from various sources
such as hand-drawn images or natural images. In this work, we concentrate on
synthesizing textures from natural images.

Traditional texture synthesis methods~\cite{wei2009state, wei2000fast,
efros1999texture, efros2001texture} aim to generate arbitrarily large texture
images indistinguishable from small input samples.
However, these approaches require holistic textures, which are rectangular and
free from geometric distortions. Obtaining such holistic textures demands
extensive human intervention~\cite{wei2009state} and is often unattainable in
most natural images. This limitation stems from frequent occlusions and
distortions in real-world objects within natural images, caused by nearby
obstructions and variations in the surface geometry of the objects. While
recent work~\cite{li2022scraping} has automated texture scraping from natural
images by grouping texture regions and filling missing regions,
it overlooks deformations, resulting in unnatural
textures. Consequently, there is a pressing need to both handle occlusions
and deformations in texture samples from real images.

We propose a novel framework that addresses these challenges by leveraging
Diffusion Models (DM)~\cite{ddpm} to synthesize holistic textures from degraded
samples in real images. Due to pixel misalignment and a lack of correspondence
between holistic textures and degraded samples, we empirically find generative
adversarial networks (GANs)~\cite{goodfellow2014generative} struggle to
synthesize holistic textures from degraded samples. We argue that GANs, due to
their susceptibility to mode-collapsing and the difficulty in capturing complex
data distributions, often produce unnatural results. DM, on the other hand,
provides a more efficient training process and produces a superior-quality of
image sample, thanks to its stationary training objective and extensive data
distribution coverage. Consequently, we adopt DM as the basis of our framework
for rectifying and synthesizing holistic textures.

Our framework builds upon Latent Diffusion Models (LDM)~\cite{rombach2022ldm}
and introduces a novel occlusion-aware latent transformer. The LDM operates in
the latent space rather than the standard pixel space, drastically reducing
computational costs. We build our framework upon LDM to allow further
downstream applications such as integration into existing photo editing tools
on personal computers. However, operating in the latent space unintentionally
entangles valid and difficult usability of the features, which originate from
occluded and unobstructed regions in sample textures. Discriminating these
features is essential for rectifying degraded textures as the invalid features
not only fail to contribute to the rectification process but can also impede
it. To address this, we introduce an occlusion-aware latent transformer into
the LDM model, delivering effective information to the rectification process.
This latent transformer utilizes partial convolutional
layers~\cite{liu2018partial} to encode the degraded sample into a latent code
composed solely of valid features, and incorporates a self-attention
block~\cite{zhang2019selfattention} to efficiently model the non-local
dependencies of the degraded sample. We empirically demonstrate the
effectiveness of the latent transformer and carefully analyze the importance of
each component.

Moreover, we introduce a method for generating synthetic training data by
applying geometric transformations and free-form mask generation to planar
textures, simulating deformations, and occlusions in degraded samples from
natural images. Specifically, we employ the homography
transformation~\cite{hartley2003multiple} and the thin plate spline
transformation~\cite{bookstein1989principal} to simulate perspective variations
and geometric distortions, respectively. We also make use of free-form
masks~\cite{yu2019freeform} to mimic occlusions found in natural images. Our
approach allows obtaining a vast number of degraded texture images from a
finite number of planar texture images by introducing varying scales of the
transformations, and enables end-to-end training of our LDM-based framework.

Experimental results attest to the superior performance of our framework
compared to existing methods. Additionally, we conduct comprehensive ablation
studies to validate the effectiveness of each component within our proposed
framework. Finally, we perform a perceptual user study that corroborates the
effectiveness of our approach.

Our contributions are summarized as follows:
\begin{itemize}[noitemsep,nolistsep,leftmargin=*]
\item The first framework for rectifying occlusions and deformations in
   degraded sample textures from natural images, expanding the applicability of
   exemplar-based texture synthesis techniques.
\rev{\item A novel occlusion-aware latent transformer that provides effective information to the texture rectification process.}
\item A synthetic data generation method to create training data for rectifying
   occlusions and deformations in texture samples. 
\item In-depth evaluation that demonstrates the superior performance of our
   framework compared to existing methods.
\end{itemize}

%% file: 02_related.tex
\section{Related Work}
In this section, we discuss the related work on texture synthesis and
generative models for image-to-image translation.

\subsection{Texture Synthesis}
Here we review the relevant literature on texture synthesis, including
exemplar-based texture synthesis, texture exemplar extraction, and shape from
texture. 

\subsubsection{Exemplar-based texture synthesis}
Exemplar-based texture synthesis aims to generate arbitrarily large new textures that are perceptually similar
to a given input sample texture. Early methods, such as
\cite{efros1999texture,efros2001texture, wei2000fast}, utilized non-parametric
techniques, copying pixels or patches sequentially while ensuring neighborhood consistency. These methods, although visually pleasing, were computationally demanding and struggled with complex patterns or large-scale structures. Recently, deep convolutional neural networks (CNNs) were employed by \cite{gatys2015texture} for texture synthesis, iterating between sample textures and random Gaussian noise. Despite revealing CNNs' potential, the optimization process remained slow. Alternative methods \cite{SGAN16, PSGAN17,li2017diversified} offered texture generation through a single CNN forward process, yet generalizing to unseen textures remained problematic. More recently,
\cite{mardani2020neural, transposer} enabled unseen texture synthesis by
upsampling textures in the Frourier domain or formulating texture synthesis as
transposed convolution operations. Nevertheless, these approaches necessitate
pristine samples, which are rectangular and free from geometric distortions.

Recently, \cite{li2022scraping} introduced an automatic texture extraction framework that groups texture regions and synthesizes large textures. Although this method can handle occlusions in degraded texture images, addressing deformations remains a significant challenge. In contrast, our framework efficiently deals with both occlusions and geometric deformations.

\subsubsection{Texture exemplar extraction}
Texture exemplar extraction is crucial in exemplar-based texture synthesis, as synthesis quality relies heavily on the selection of representative texture samples. Traditionally, this process is labor-intensive, necessitating expert input and significant resources. To mitigate this, \cite{wu2018automatic} introduced an automated method for extracting texture exemplars from images, utilizing both global and local textureness measures. Building on this, \cite{wu2021deep} proposed a deep learning-based approach for texture exemplar extraction. However, frequent occlusions and deformations in natural images can hinder the extraction of appropriate exemplars. Therefore, rectifying these occlusions and distortions, as proposed in our framework, is key to improving the texture synthesis process. 

\subsubsection{Shape from texture}
Shape from texture, a subfield of computer vision and image processing, focuses on deriving 3D shape information from 2D images or textures. The goal is to extract depth, orientation, and other geometric properties from the arrangement of texture elements, allowing for the reconstruction of planar textures. Representative work by \cite{verbin2020toward} formulates the problem as a three-player game to convert an input image into a 2.5D shape and a planar texture. However, while capable of estimating depth and creating planar textures, shape-from-texture methods struggle with structured textures and require significant computational time per input image. 

In summary, despite the advancements in texture synthesis, synthesizing textures from natural images is still challenging due to occlusions and deformations. Our framework addresses these issues by rectifying these elements in sample textures, ultimately enhancing the performance and applicability of texture synthesis from natural images.

\subsection{Generative Models for Image-to-image Translation}
We tackle the rectification of occlusions and deformations as an image-to-image
translation problem, a process that converts an input image from one domain to a corresponding image in another, while preserving crucial structural and contextual details. In our task, we consider converting a degraded texture sample into a planar texture, maintaining the overall texture appearance and structure. Therefore, we delve into several recent generative models for the image-to-image translation problem. 

\subsubsection{Generative Adversarial Networks (GANs)}
GANs~\cite{goodfellow2014generative} consist of a generator and a discriminator
that play an adversarial game to generate realistic samples from a prior
distribution. GANs have been extensively employed in image-to-image translation
tasks, with notable examples being \cite{isola2017image}, which uses a
conditional GAN~\cite{cGAN} to learn a mapping between input and output images,
and \cite{zhu2017unpaired}, which extends this concept to unpaired image
translation.
Several existing approaches~\cite{nonstat,transposer,mardani2020neural} in
exemplar-based texture synthesis have also adopted GANs as a basis to generate
textures.
We find that GANs, due to their susceptibility to the
mode-collapse problem and challenges in capturing complex data distributions,
often produce unnatural results in difficult tasks like ours.

\subsubsection{Diffusion probabilistic models}
Recently, diffusion probabilistic models (DM)~\cite{sohl2015deep} have taken
the lead in the image synthesis field in terms of both sample quality and
diversity. \cite{ddpm} presented
Denoising Diffusion Probabilistic Model (DDPM) for high-quality image synthesis and achieved
sample quality comparable to GANs. \cite{song2021denoising, song2021score} exploited advances in score-based generative modeling for accurate score estimation and efficient sample generation. A seminal work~\cite{guided_diffusion} showcased that DM can attain superior image sample quality compared to GANs. With the advent of classifier-free guidance~\cite{ho2021classifierfree}, the necessity of an external classifier in the generation process of conditional DM was eliminated. 

Various applications using DM have since emerged. For instance, \cite{sr3} utilized DMs for conditional image generation, achieving superior performance in various super-resolution tasks and producing more realistic outputs than GAN-based methods. Notably, \cite{saharia2022palette} introduced a unified framework for image-to-image translation using conditional DM, paving the way for DM in image-to-image translation tasks. However, the use of DM has been limited by its extensive computational resource demands during both training and sampling. This not only impedes progress in the field but also constrains downstream applications. To mitigate this limitation, \cite{rombach2022ldm} proposed latent diffusion models (LDM) to reduce computational resources for DM while maintaining their quality and flexibility. By training DM on the latent representation of a pre-trained vector-quantized variational autoencoder (VQ-VAE)~\cite{vandenOord2017neural}, LDM achieves competitive results in various tasks with reduced computational costs. Our framework builds upon the LDM for further downstream applications such as integration into existing photo editing tools on personal computers. 

%% file: 03_method.tex
\begin{figure*}[t]
  \centering
  \includegraphics[width=\textwidth]{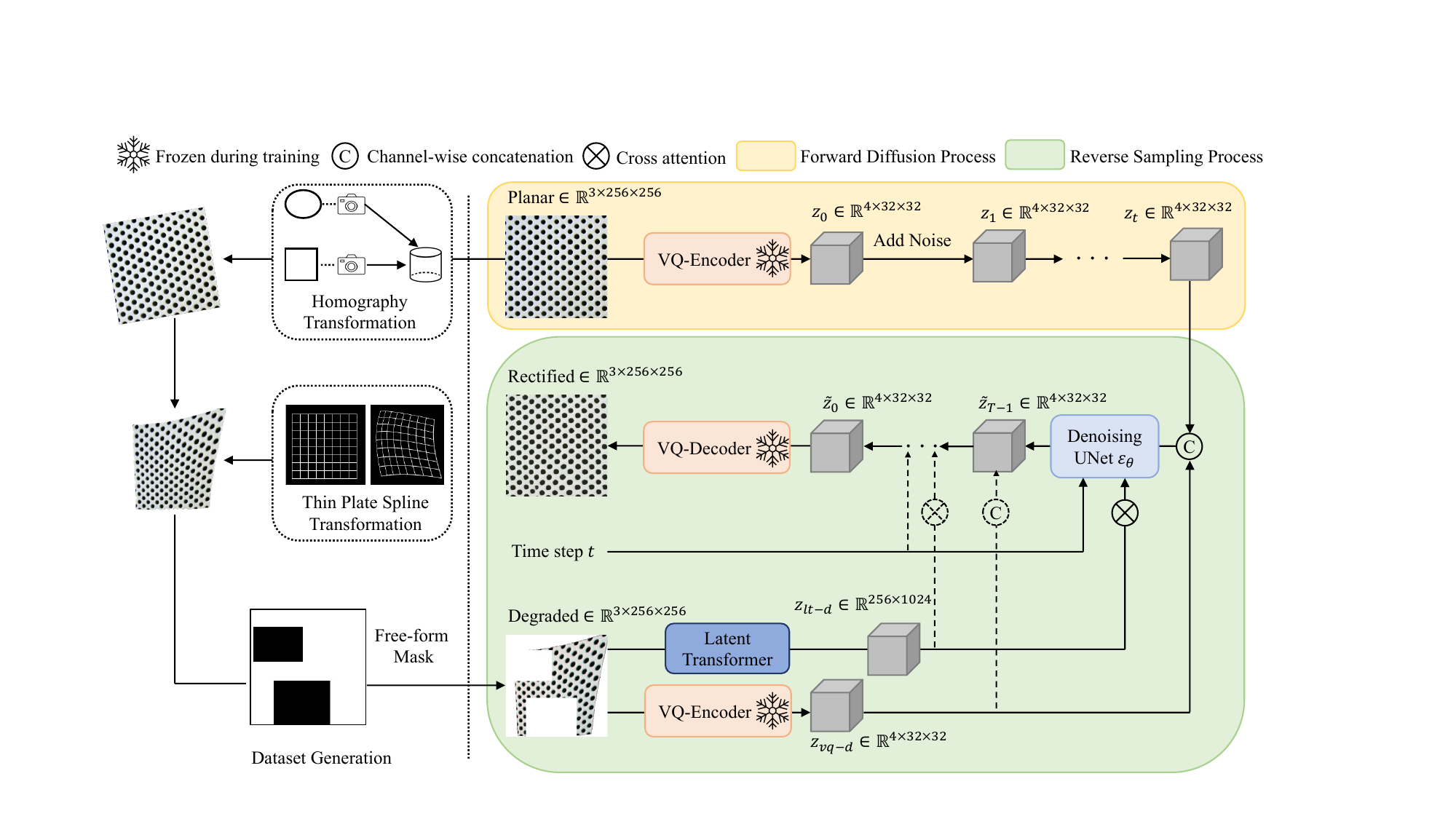}
  \caption{\textbf{An overview of the proposed framework.} Our synthetic
  training dataset is constructed by applying random geometric transformations and
  free-form masks on planar textures.
  During the training phase, our framework takes as input both degraded and
  planar textures, and performs forward diffusion and reverse sampling
  processes. Upon completion of the training, our approach takes as input a
  degraded texture sample and outputs a rectified texture.}
  \label{fig:framework}
\end{figure*}

\section{Approach}
In this section, we introduce our proposed framework for rectifying occlusions
and deformations in degraded sample textures. Our framework is based on a
Latent Diffusion Model~\cite{rombach2022ldm}. We enable conditional generation
by concatenating a latent code of the degraded sample with random noise, while
also incorporating features from an occlusion-aware latent transformer using
cross-attention layers. An overview of the framework is depicted in
Fig.~\ref{fig:framework}. 

\subsection{Preliminary: Latent Diffusion Models (LDM)}
\label{sec:prelim}
\rev{Our framework for rectifying deformations and occlusions is built on the LDM~\cite{rombach2022ldm}. This allows for integration with existing photo editing tools due to its lower memory consumption compared to pixel-based DMs. 
The LDM utilizes a VQ-VAE~\cite{vandenOord2017neural} encoder $\mathcal{E}_{vq}$ to encode a planar texture $\mathbf{P} \in \mathbb{R}^{C\times H\times W}$ into a latent code $z_{0} \in \mathbb{R}^{c\times h\times w}$. During training, the forward diffusion process $FwdDiff$ incrementally introduces Gaussian noise to the latent code $z_{0}$ at timestep $t\sim \mathcal{U}(1, T)$,
and the reverse denoising process subsequently denoises a corrupted latent code $z_{t}$ at timestep $t$ using a trainable denoising network $\epsilon_{\theta}$.
After training, we
acquire a trained denoising network $\epsilon_{\theta}$ that predicts the noise added at
timestep $t$ given $z_{t}$. With the trained denoising network, we can synthesize
a planar texture from scratch by iteratively denoising on a Gaussian noise $z_{T} \sim
\mathcal{N}(0, \mathbf{I})$.  
}

\subsection{Conditional Generation}
\label{sec:condit}
In addition to the above unconditional generation, we employ conditioning
mechanisms to constrain the generated textures on degraded samples. Following
the concept of classifier-free diffusion guidance~\cite{ho2021classifierfree},
we train our model by modeling the conditional distributions $p(\mathbf{P} \mid
\mathbf{D})$, where $\mathbf{P}$ and $\mathbf{D}$ are planar textures and
degraded textures, respectively. More specifically, we incorporate concatenation and
cross-attention conditioning mechanisms into our framework to ensure the textures generated correspond to the degraded textures. 

In the concatenation mechanism, the diffused latent code $\tilde{z}_{t}$ is paired with the latent code $z_{vq-d}$ of a degraded sample. Both latent codes, encoded using the same VQ-VAE encoder, are concatenated along the channel dimension.
Although this concatenation ensures that the identity of the degraded samples
is maintained, the latent code $z_{vq-d}$ unintentionally entangles valid and
invalid features coming from occluded and valid regions in the degraded sample.
This can mislead the reverse denoising process and produce unnatural results.

\rev{We address the entanglement of the latent code $z_{vq-d}$ by introducing an occlusion-aware latent transformer $\tau_{\theta}$. This transformer is trained from scratch to offer valid guidance during the generation process. We integrate this guidance into the generation process using cross-attention layers.}
Let $z_{lt-d} \in \mathbb{R}^{C_{lt} \times d_{lt}}$ be
compensatory feature obtained by the occlusion-aware latent transformer
$\tau_{\theta}$, and $\varphi_i\left(\tilde{z}_t\right) \in
\mathbb{R}^{C_{inter}^{i} \times d_{inter}^{i}}$ be flattened intermediate
features before the $i$-th cross-attention layer. Conditional generation with
cross-attention layers is implemented as:
\begin{gather}
    Q = W_Q^{(i)} \cdot \varphi_i\left(\tilde{z}_t\right), \quad
     K = W_K^{(i)} \cdot z_{lt-d}, \quad
     V = W_V^{(i)} \cdot z_{lt-d}, \nonumber \\
    Attention(Q, K, V) = softmax(\left(\frac{Q K^T}{\sqrt{d}}\right) \cdot V),
  \label{eq:crossattn}
\end{gather}
\noindent where $W_Q^{(i)}$, $W_K^{(i)}$, and $W_V^{(i)}$ are learnable
encoding functions. 

The final training objective used for training the conditional LDM with pairs
of planar textures and degraded samples $\left \{(\mathit{\mathbf{P_{i}}},
\mathit{\mathbf{D_{i}}})\right \}_{i=1}^{K}$ can be formally described as
follows:
\begin{gather}
  z_{vq-d} = \mathcal{E}_{vq}(\mathbf{D}), \quad
  z_{lt-d}  = \tau_{\theta}(\mathbf{D}), \quad
  \tilde{z}_{t}  = FwdDiff(\mathcal{E}_{vq}(\mathbf{P})), \nonumber \\
  L_{L D M}=\mathbb{E}_{\epsilon \sim \mathcal{N}(0,1), t \sim \mathcal{U}(1, T)}\left[\left\|\epsilon-\epsilon_\theta\left(\tilde{z}_t, t,z_{vq-d}, z_{lt-d}\right)\right\|^2\right],
\end{gather}
\noindent where $FwdDiff$ refers to the forward diffusion process that adds noise according to the scheduler. \rev{Note that the $FwdDiff$ is only used during training and is replaced with a random Gaussian noise $z_{T} \sim \mathcal{N}(0, \mathbf{I})$ during inference. }
With these conditioning mechanisms and conditional training objectives, our framework is able to map the degraded textures to planar textures.

\subsection{Occlusion-aware Latent Transformer}
\label{sec:lt}
Since the entangled features $z_{vq-d}$, mislead the generation process, we
propose to use a novel occlusion-aware latent transformer to compensate for the
entangled features. This latent transformer takes as input a degraded texture
and outputs valid guidance to the generation process while capturing
long-range dependencies.
We achieve these capacities with two key components: partial
convolutional layers for occlusion elimination and self-attention block for
modeling long-range relationships. 
\rev{Table~\ref{lt_specs} provides full details of our occlusion-aware latent transformer architecture.}

\begin{table}[tb]
   \caption{\rev{\textbf{Architecture of the occlusion-aware latent transformer.} 
   The input layer takes as input a concatenation of a sample texture and its corresponding mask. Each \textit{PartialConv} layer consists of a sequence: a partial convolution layer, followed by a Batch Norm layer, and then a \textit{ReLU} layer. At the end of the latent transformer, the output feature is flattened to a size of $256 \times 1024$.
   }}
   \centering
   \begin{tabular}{cccc}
   \toprule
   Layer Type & Kernel & Strides & Output Resolution \\
   \midrule
   Input\&Mask & - & - & $6 \times 256 \times 256$ \\
   PartialConv & $3 \times 3$ & $2 \times 2$ & $64 \times 128 \times 128$ \\
   PartialConv & $3 \times 3$ & $1 \times 1$ & $128 \times 128 \times 128$ \\
   PartialConv & $3 \times 3$ & $2 \times 2$ & $128 \times 64 \times 64$ \\
   PartialConv & $3 \times 3$ & $1 \times 1$ & $256 \times 64 \times 64$ \\
   PartialConv & $3 \times 3$ & $2 \times 2$ & $256 \times 32 \times 32$ \\
   PartialConv & $3 \times 3$ & $1 \times 1$ & $512 \times 32 \times 32$ \\
   PartialConv & $3 \times 3$ & $1 \times 1$ & $512 \times 32 \times 32$ \\
   Self-attention & $3 \times 3$ & $1 \times 1$ & $512 \times 32 \times 32$ \\
   PartialConv & $3 \times 3$ & $1 \times 1$ & $256 \times 32 \times 32$ \\
   Flatten layer & $3 \times 3$ & $1 \times 1$ & $256 \times 1024 $ \\
   \bottomrule
   \end{tabular}
   \label{lt_specs}
   \end{table}

\subsubsection{Occlusion Elimination}
Distilling valid features from degraded samples is important as these samples
often carry invalid information stemming from occlusions and geometric
deformations. Building on the concept introduced in \cite{liu2018partial}, we
employ partial convolutional layers for the extraction of valid features from
these samples, effectively mitigating the effects of occlusions. 
In detail, given input features $x$ and a corresponding
mask $m$, in which 0 and 1 indicate invalid and valid regions respectively, the
partial convolutional layer is defined as:
\begin{equation}
  x^{\prime}= \begin{cases}
   \mathbf{W}^T(x \odot m) \frac{\operatorname{sum}(\mathbf{1})}{\operatorname{sum}(m)}+b, & \text { if } \operatorname{sum}(m)>0 \\
   0, & \text { otherwise }
  \end{cases}, 
\end{equation}
\noindent where $\odot$ denotes Hadamard product, while $\mathbf{W}$ and $b$
represent the convolution filters and the corresponding bias, respectively. The
input feature $x$ can either be degraded texture or any intermediate feature. After each partial convolutional layer, the current mask $m$ is updated with the following definition:
\begin{equation}
  m^{\prime}=\begin{cases}
   1, & \text { if } \operatorname{sum}(m)>0 \\
   0, & \text {otherwise}
  \end{cases}.
\end{equation}

We apply the partial convolutional layer sequentially eight times, where
the downsampling operation is performed three times. By repeatedly applying the
layer with downsampling operations, we eventually obtain a valid feature in
latent representation. This latent representation subsequently offers valid
guidance to the texture rectification process. 

\subsubsection{Modeling Long-Range Dependencies}
While the partial convolutional layers are proficient in addressing occlusions,
they fall short in modeling long-range dependencies.
\rev{
Modeling long-range dependencies is crucial for rectifying degraded textures, especially since valid information in these textures is often sparse due to occlusions. To address this, we incorporate a self-attention layer~\cite{zhang2019selfattention} at the end of the latent transformer. This allows for the calculation of non-local relationships from sparse information, thereby capturing long-range contextual information.
}
This self-attention layer, \rev{which can be construed as a variant of the cross-attention layer (Eq.~\ref{eq:crossattn}) with a single input feature, }
can then generate an output
feature that guides the texture rectification process through the subsequent
cross-attention layers. 

\rev{Overall, our proposed occlusion-aware latent transformer addresses occlusions using partial convolution layers and captures long-range dependencies through the self-attention layer. The resulting valid guidance is then integrated into the texture rectification process via the cross-attention conditioning mechanism, leading to enhanced performance in the texture rectification and synthesis task.
}

\section{Dataset}
We generate synthetic training data by applying homography
transformation~\cite{hartley2003multiple}, thin plate spline
transformation~\cite{bookstein1989principal}, and free-form
mask~\cite{yu2019freeform} to planar textures, simulating
perspective variations, geometric deformations, and occlusions. We first collect texture images from multiple
sources~\cite{bell13opensurfaces,safia2013new,burghouts2009material,cimpoi14describing,dai2014synthesizability,sharan2014accuracy,salzburg_texture_db,Mallikarjuna2006THEK2,vistex_texture} and manually filter out images that already exhibit degradations. After filtering, we obtain
a collection of 22,043 planar texture images. And then, we perform homography
transformation and thin plate spline (TPS) transformation on these planar
texture images to simulate perspective variations and geometric distortions. Subsequently, free-form masks are applied to mimic occlusions. A visual
illustration of the synthetic data generation is shown in
Fig.~\ref{fig:framework}. 

We incorporate randomness into the generation process by varying the scale of transformations. Specifically, homography transformation is applied with a distortion scale $ s_{hmg} \sim
\mathcal{U}(0.3, 0.5)$ and a probability of 80\%, while the TPS transformation is employed with a distortion scale $ s_{tps} \sim \mathcal{U}(0.1, 0.3)$ and a probability of 80\%. These transformations are implemented using the Kornia library~\cite{eriba2019kornia}. This random generation process produces a diverse set of degraded texture images from a finite pool of planar textures, which is crucial for learning holistic texture rectification and synthesis in an end-to-end manner. The synthetic dataset is split into a training set with 15,430 images, a validation set with 2,205 images, and a test set with 4,408 images. 


%% file: 04_exps.tex
\section{Experimental Results}
In this section, we present a comprehensive evaluation of our texture
rectification framework.

\subsection{Implementation Details}
Our framework is trained for one million iterations on the proposed dataset
with a batch size of 32. This takes approximately 4 days on eight A100 GPUs. The sampling process is performed on a single
RTX 3090 GPU. The input patch used during the training phase is first resized
to 294$\times$294 pixels and then randomly cropped 256$\times$256 pixels from
the resized one. We employ the Adam optimizer~\cite{kingma2015adam} with a learning rate of 1e-6. The diffusion process operates with a linear noise schedule, ranging from 0.0015 to
0.0195, which is distributed over 1000 time steps. For sampling, we
utilize 200 steps of the Denoising Diffusion Implicit Model (DDIM) strategy~\cite{song2021denoising}, which requires 4 seconds to generate a 256$\times$256 texture.

\subsection{Evaluation Metrics}
Following the common metrics used in the field of texture
synthesis~\cite{transposer, li2022scraping}, we employ a set of evaluation
metrics that captures various aspects of texture images to assess the
\rev{occlusion elimination and geometric rectification  capabilities of our framework.} 
Specifically, we use the following metrics to assess the content preservation, reconstruction quality, style consistency, and distribution match between generated and real planar textures:
\begin{itemize}[noitemsep,nolistsep,leftmargin=*]
\item \textbf{Structural Similarity Index Measure (SSIM):} The
   SSIM~\cite{wang2004image} measures the preservation of structural information
   in the rectified textures with a larger value indicating higher similarity.
\item \textbf{Learned Perceptual Image Patch Similarity (LPIPS):} The
   LPIPS~\cite{zhang2018unreasonable} quantifies perceptual differences between
   images with a lower score indicating higher similarity.
\item \textbf{Gram Matrix Distance (GMD): } We use the
   GMD~\cite{jognson_perceptual_2016} to evaluate the style consistency with a
   lower score indicating closer matching in texture style.
\item \textbf{Fréchet Inception Distance (FID):} We employ the
   FID~\cite{heusel17} to measure the statistical similarity between
   distributions of images with a lower score indicating higher similarity.
\end{itemize}

\subsection{Baselines}
Our task inherently relates to the problems of image-to-image translation as we
consider converting degraded texture to planar texture. Among these problems, image inpainting is closely related to our task as it also
handles occlusions. We compare our approach against several representative
methods recognized for their performance in these areas to provide a
comprehensive evaluation of our approach. These comparison baselines include
pix2pix~\cite{isola2017image} and VQGAN~\cite{esser2021taming}, well-known for image-to-image translation methods, and a leading
method in image inpainting.
All approaches are trained on the same dataset as our approach. The implementation details of the comparisons can be found in the supplemental. 

\begin{itemize}[noitemsep,nolistsep,leftmargin=*]
\item \textbf{pix2pix:} A widely-adopted Generative Adversarial Network-based
   image-to-image translation framework~\cite{isola2017image}.

\item \textbf{VQGAN:} The Vector Quantized Generative Adversarial Network
   (VQGAN)~\cite{esser2021taming} represents the state-of-the-art for diverse
   image-to-image translation tasks.

\item \textbf{MAT:} Given the inpainting aspect of our task, we draw a
   comparison with a leading transformer-based image inpainting
   method~\cite{li2022mat}. 
\end{itemize}

\begin{table}[t]
   \centering
   \caption{\textbf{Quantitative results.} Comparative analysis of our method
   against other texture generation models, considering different metrics:
   SSIM, LPIPS, GMD, and FID. The table presents results that highlight the
   superiority of our method in terms of these metrics. For a fair comparison,
   all methods were trained on the synthetic training dataset and evaluated on
   the synthetic test dataset. }
   \label{table:quan_eval}
   \begin{tabular}{rcccc}
   \toprule
   Method  & SSIM $\uparrow$& LPIPS $\downarrow$& GMD $\downarrow$& FID $\downarrow$\\
   \midrule
   pix2pix & 0.0141 & 0.7742 & 39.29 & 607.15 \\
   MAT & 0.2466 & 0.6751 & 34.17 & 187.40  \\
   VQGAN & 0.4549 & 0.4407 & 24.65 & 45.21 \\
   Ours & \textbf{0.5096} & \textbf{0.3417} & \textbf{15.32} & \textbf{15.50}  \\
   \bottomrule
   \end{tabular}
\end{table}

\subsection{Quantitative Evaluation}
We assess the performance of our method against baselines pix2pix,
VQGAN, and MAT using the LPIPS, SSIM, GMD, and FID metrics. As demonstrated in Table~\ref{table:quan_eval}, our method consistently outperforms the others. Key aspects contributing to these quantitative results include:

\paragraph{Occlusion and Deformation Handling:}
Our framework effectively addresses occlusions and deformations, as indicated by the low LPIPS and high SSIM scores. This suggests our method generates textures that are perceptually and structurally more similar to the planar textures compared to the other methods. 

\paragraph{Texture Preservation:}
The lower GMD of our method signifies a higher degree of texture feature preservation, demonstrating the capability of our framework in effectively extracting valid features from degraded textures. 

\paragraph{Quality of Generated Images:}
The FID scores suggest our synthesized textures match the statistical properties of planar texture more
closely than other methods, indicating the effectiveness in accurately learning the distributions of planar textures. 

In summary, the superior performance of our framework in these quantitative
evaluations underscores its effectiveness in holistic texture rectification and synthesis.
The qualitative evaluations in the next section provide further visual evidence
to support these findings.

\subsection{Qualitative Evaluation}
We provide visual results of our method alongside the outputs of VQGAN and MAT baselines. As shown in Figure~\ref{fig:synthetic_comp}, our method consistently generates visually superior results, effectively handling occlusions and distortions, reconstructing detailed texture information, and producing results perceptually closer to the planar textures. We also validate the effectiveness of our method on real-world textures, feeding selected regions from real images to our framework and the baseline methods. As evident in Figure~\ref{fig:real_comp}, our method successfully preserves the texture and overall structure of the selected regions, delivering visually pleasing results that remain perceptually closer to the original textures.  

In contrast to our framework that synthesizes realistic results, other methods often yield unnatural textures. Although MAT can produce texture-like images, it succumbs to the mode-collapse problem, resulting in outputs that disregard the degraded textures. VQGAN, despite effectively capturing data distributions, generates incorrect results relative to the degraded
textures. We exclude the results of pix2pix in visual comparison as it persistently produces all-black images. 

\begin{table}[tbp]
    \centering
    \caption{\textbf{Results of the perceptual user study.} The table presents the percentage of times each method was preferred over the others for generating more realistic textures, as determined by human evaluators. A higher percentage indicates a method was often favored due to its superior quality in texture rectification.}
    \begin{tabular}{rccc}
    \toprule
       & MAT & VQGAN & Ours \\
     \midrule
    vs. MAT & - & 85.40\% & \textbf{90.56\%} \\
    vs. VQGAN  & 14.60\% & - & \textbf{75.32\%} \\
    vs. Ours & 9.44\% & 24.68\%& - \\
     \bottomrule
    \end{tabular}
    \label{tab:usestudy}
\end{table}

\subsection{User Study}
To further validate the perceptual quality of our synthesized
results, we conduct a user study with synthetic data and real images. The goal is to get a human perspective on the effectiveness of our method compared to baselines. We first use synthetic test images for the study, considering the potential complexity of our task for laypersons. We randomly select 50 images from the test set for participants, showing them a degraded texture and a ground truth image. Participants are asked to select the more realistic image from two randomly generated results by different methods. After ensuring participants understand our task through this initial study with synthetic data, we conduct a user study with real images. We prepare 63 real images, manually selected a desired texture region for each, and generate synthesized textures using each of the three methods. We randomly pick 50 images from these 63 images and ask participants to select the more realistic one from two randomly chosen results. \rev{14 laypersons participate in the study, and each of them provides} 50 sets of feedback on synthetic test images and 50 sets on real images. As shown in Table~\ref{tab:usestudy}, our method is preferred 90.56\% of the time compared to MAT and 75.32\% compared to VQGAN. This user study underscores the perceptual superiority of our framework in holistic texture rectification and synthesis, as it is consistently favored over the baselines.

\begin{table}[htbp]
    \centering
    \caption{\textbf{Ablation study on the occlusion-aware latent transformer and conditioning mechanism.} The table presents a performance comparison of different configurations of our framework, including the influence of the self-attention layer and partial convolutional layers in the occlusion-aware latent transformer, and the impact of various conditioning mechanisms. The `Conditions' and `Arch.' represent the conditioning mechanism and architecture of the latent transformer. \rev{The terms `Concatenation' and `Crossattn' refer to the `only concatenation' and `only cross-attention' conditioning mechanisms, respectively.} The `PCE' represents removing the self-attention block from the latent transformer, and the `SAE' indicates replacing the partial convolutions with standard convolutions.}
    \begin{tabular}{rccccc}
    \toprule
    Conditions & Arch. & SSIM $\uparrow$& LPIPS $\downarrow$& GMD $\downarrow$& FID $\downarrow$ \\
    \midrule
    Full & Full & \textbf{0.5096} & \textbf{0.3417} & \textbf{15.32} & \textbf{15.50} \\
    Concatenation & Full & 0.4842 & 0.3621 & 16.37 & 17.68 \\
    Crossattn & Full & 0.4721 & 0.3857 & 23.47 & 21.77 \\
    Full & SAE & 0.4865 & 0.3608 & 15.78 & 17.85 \\
    Full & PCE & 0.4879 & 0.3596 & 17.61 & 16.15 \\
    \bottomrule
    \end{tabular}
    \label{tab:ablation}
\end{table}

\subsection{Ablation Study}
In order to further evaluate the effectiveness of various components of our
method, we conduct an ablation study that 
the occlusion-aware latent transformer and conditioning mechanisms. 

\subsubsection{Conditioning Mechanism}
We explore the conditioning mechanisms of our framework, comparing three different configurations: only concatenation \rev{with the latent code $z_{vq-d}$}, only cross-attention \rev{with the compensatory feature $z_{lt-d}$}, and our full method combining both. As Table~\ref{tab:ablation} and Figure~\ref{fig:ab_conds} show, the full method yields the best performance overall. These results validate our hypothesis that while concatenation provides overall guidance, cross-attention offers essential valid features. 

\subsubsection{Occlusion-Aware Transformer}
We also conduct an ablation study on each component of our occlusion-aware
latent transformer to evaluate their contributions to the overall performance.
In particular, we evaluate the impact of removing the self-attention layer and replacing partial convolutional layers with standard ones. Table~\ref{tab:ablation} presents the results, and Fig.~\ref{fig:ab_lt} provides visual outcomes. The results show that both the partial convolutional layer and the self-attention layer proved crucial to holistic texture rectification and synthesis. Their removal or replacement led to significant performance reduction, underscoring the importance of these components in effectively rectifying degraded textures. 

%% file: 05_conc.tex
\section{Limitations and Discussion}
Despite its effectiveness, our method has limitations, most notably the requirement for fixed-size degraded textures, limiting the flexibility and usage scenarios of our approach. Future work should focus on addressing this limitation. Recent advancements~\cite{bar-tal2023multidiffusion} in generating arbitrarily large images with DM present potential solutions for varying input sizes. Such improvements could broaden the applicability of our approach, making it more versatile for tasks such as synthesizing large textures from degraded samples. This advancement is expected to substantially contribute to texture synthesis research. 
\rev{Additionally, our framework occasionally produces imperfect results, particularly in cases with varying lighting conditions and extreme distortions in the sample images. We can address these issues by masking regions with significant lighting changes and incorporating more training data with extreme distortions.}

%% file: top.bbl

\begin{thebibliography}{55}


\ifx \showCODEN    \undefined \def \showCODEN     #1{\unskip}     \fi
\ifx \showDOI      \undefined \def \showDOI       #1{#1}\fi
\ifx \showISBNx    \undefined \def \showISBNx     #1{\unskip}     \fi
\ifx \showISBNxiii \undefined \def \showISBNxiii  #1{\unskip}     \fi
\ifx \showISSN     \undefined \def \showISSN      #1{\unskip}     \fi
\ifx \showLCCN     \undefined \def \showLCCN      #1{\unskip}     \fi
\ifx \shownote     \undefined \def \shownote      #1{#1}          \fi
\ifx \showarticletitle \undefined \def \showarticletitle #1{#1}   \fi
\ifx \showURL      \undefined \def \showURL       {\relax}        \fi
\providecommand\bibfield[2]{#2}
\providecommand\bibinfo[2]{#2}
\providecommand\natexlab[1]{#1}
\providecommand\showeprint[2][]{arXiv:#2}

\bibitem[Abdelmounaime and Dong-Chen(2013)]%
        {safia2013new}
\bibfield{author}{\bibinfo{person}{Safia Abdelmounaime} {and}
  \bibinfo{person}{He Dong-Chen}.} \bibinfo{year}{2013}\natexlab{}.
\newblock \showarticletitle{New Brodatz-Based Image Databases for Grayscale
  Color and Multiband Texture Analysis}.
\newblock \bibinfo{journal}{\emph{Volume 2013}} (\bibinfo{year}{2013}).
\newblock
\urldef\tempurl%
\url{https://doi.org/10.1155/2013/876386}
\showDOI{\tempurl}


\bibitem[Bar-Tal et~al\mbox{.}(2023)]%
        {bar-tal2023multidiffusion}
\bibfield{author}{\bibinfo{person}{Omer Bar-Tal}, \bibinfo{person}{Lior Yariv},
  \bibinfo{person}{Yaron Lipman}, {and} \bibinfo{person}{Tali Dekel}.}
  \bibinfo{year}{2023}\natexlab{}.
\newblock \showarticletitle{MultiDiffusion: Fusing Diffusion Paths for
  Controlled Image Generation}. In \bibinfo{booktitle}{\emph{International
  Conference on Machine Learning}}.
\newblock


\bibitem[Bell et~al\mbox{.}(2013)]%
        {bell13opensurfaces}
\bibfield{author}{\bibinfo{person}{Sean Bell}, \bibinfo{person}{Paul Upchurch},
  \bibinfo{person}{Noah Snavely}, {and} \bibinfo{person}{Kavita Bala}.}
  \bibinfo{year}{2013}\natexlab{}.
\newblock \showarticletitle{Open{S}urfaces: A Richly Annotated Catalog of
  Surface Appearance}.
\newblock \bibinfo{journal}{\emph{ACM Transactions on Graphics (Proceedings of
  SIGGRAPH)}} \bibinfo{volume}{32}, \bibinfo{number}{4} (\bibinfo{year}{2013}).
\newblock


\bibitem[Bergmann et~al\mbox{.}(2017)]%
        {PSGAN17}
\bibfield{author}{\bibinfo{person}{Urs Bergmann}, \bibinfo{person}{Nikolay
  Jetchev}, {and} \bibinfo{person}{Roland Vollgraf}.}
  \bibinfo{year}{2017}\natexlab{}.
\newblock \showarticletitle{Learning Texture Manifolds with the Periodic
  Spatial GAN}. In \bibinfo{booktitle}{\emph{International Conference on
  Machine Learning}}.
\newblock


\bibitem[Bookstein(1989)]%
        {bookstein1989principal}
\bibfield{author}{\bibinfo{person}{Fred~L Bookstein}.}
  \bibinfo{year}{1989}\natexlab{}.
\newblock \showarticletitle{Principal warps: Thin-plate splines and the
  decomposition of deformations}.
\newblock \bibinfo{journal}{\emph{IEEE Transactions on Pattern Analysis and
  Machine Intelligence}} \bibinfo{volume}{11}, \bibinfo{number}{6}
  (\bibinfo{year}{1989}), \bibinfo{pages}{567--585}.
\newblock


\bibitem[Burghouts and Geusebroek(2009)]%
        {burghouts2009material}
\bibfield{author}{\bibinfo{person}{Gertjan~J. Burghouts} {and}
  \bibinfo{person}{Jan-Mark Geusebroek}.} \bibinfo{year}{2009}\natexlab{}.
\newblock \showarticletitle{Material-specific Adaptation of Color Invariant
  Features}.
\newblock \bibinfo{journal}{\emph{Pattern Recognition Letters}}
  \bibinfo{volume}{30} (\bibinfo{year}{2009}), \bibinfo{pages}{306--313}.
\newblock


\bibitem[Chen and Rosenberg(2018)]%
        {chen2018virtual}
\bibfield{author}{\bibinfo{person}{Chin-Fan Chen} {and}
  \bibinfo{person}{Evan~Suma Rosenberg}.} \bibinfo{year}{2018}\natexlab{}.
\newblock \showarticletitle{Virtual Content Creation Using Dynamic
  Omnidirectional Texture Synthesis}. In \bibinfo{booktitle}{\emph{IEEE
  Conference on Virtual Reality and 3D User Interfaces (VR)}}.
\newblock


\bibitem[Cimpoi et~al\mbox{.}(2014)]%
        {cimpoi14describing}
\bibfield{author}{\bibinfo{person}{Mircea Cimpoi}, \bibinfo{person}{Subhransu
  Maji}, \bibinfo{person}{Iasonas Kokkinos}, \bibinfo{person}{Sammy Mohamed},
  {and} \bibinfo{person}{Andrea Vedaldi}.} \bibinfo{year}{2014}\natexlab{}.
\newblock \showarticletitle{Describing Textures in the Wild}. In
  \bibinfo{booktitle}{\emph{IEEE Conference on Computer Vision and Pattern
  Recognition}}.
\newblock


\bibitem[Criminisi et~al\mbox{.}(2003)]%
        {criminisi2003object}
\bibfield{author}{\bibinfo{person}{Antonio Criminisi}, \bibinfo{person}{Patrick
  Pérez}, {and} \bibinfo{person}{Kentaro Toyama}.}
  \bibinfo{year}{2003}\natexlab{}.
\newblock \showarticletitle{Object Removal by exemplar-based inpainting}. In
  \bibinfo{booktitle}{\emph{IEEE Conference on Computer Vision and Pattern
  Recognition}}.
\newblock


\bibitem[Dai et~al\mbox{.}(2014)]%
        {dai2014synthesizability}
\bibfield{author}{\bibinfo{person}{Dengxin Dai}, \bibinfo{person}{Hayko
  Riemenschneider}, {and} \bibinfo{person}{Luc Van~Gool}.}
  \bibinfo{year}{2014}\natexlab{}.
\newblock \showarticletitle{The Synthesizability of Texture Examples}. In
  \bibinfo{booktitle}{\emph{IEEE Conference on Computer Vision and Pattern
  Recognition}}.
\newblock


\bibitem[Dhariwal and Nichol(2021)]%
        {guided_diffusion}
\bibfield{author}{\bibinfo{person}{Prafulla Dhariwal} {and}
  \bibinfo{person}{Alexander Nichol}.} \bibinfo{year}{2021}\natexlab{}.
\newblock \showarticletitle{Diffusion Models Beat GANs on Image Synthesis}. In
  \bibinfo{booktitle}{\emph{Advances in Neural Information Processing
  Systems}}.
\newblock


\bibitem[Efros and Freeman(2001)]%
        {efros2001texture}
\bibfield{author}{\bibinfo{person}{Alexei~A. Efros} {and}
  \bibinfo{person}{William~T. Freeman}.} \bibinfo{year}{2001}\natexlab{}.
\newblock \showarticletitle{Image Quilting for Texture Synthesis and Transfer}.
  In \bibinfo{booktitle}{\emph{SIGGRAPH '01: Proceedings of the 28th Annual
  Conference on Computer Graphics and Interactive Techniques}}.
  \bibinfo{pages}{341–346}.
\newblock


\bibitem[Efros and Leung(1999)]%
        {efros1999texture}
\bibfield{author}{\bibinfo{person}{Alexei~A. Efros} {and}
  \bibinfo{person}{Thomas~K. Leung}.} \bibinfo{year}{1999}\natexlab{}.
\newblock \showarticletitle{Texture Synthesis by Non-parametric Sampling}. In
  \bibinfo{booktitle}{\emph{International Conference on Computer Vision}}.
\newblock


\bibitem[Esser et~al\mbox{.}(2021)]%
        {esser2021taming}
\bibfield{author}{\bibinfo{person}{Patrick Esser}, \bibinfo{person}{Robin
  Rombach}, {and} \bibinfo{person}{Björn Ommer}.}
  \bibinfo{year}{2021}\natexlab{}.
\newblock \showarticletitle{Taming Transformers for High-Resolution Image
  Synthesis}. In \bibinfo{booktitle}{\emph{IEEE Conference on Computer Vision
  and Pattern Recognition}}.
\newblock


\bibitem[Gatys et~al\mbox{.}(2015)]%
        {gatys2015texture}
\bibfield{author}{\bibinfo{person}{Leon Gatys}, \bibinfo{person}{Alexander~S.
  Ecker}, {and} \bibinfo{person}{Matthias Bethge}.}
  \bibinfo{year}{2015}\natexlab{}.
\newblock \showarticletitle{Texture Synthesis Using Convolutional Neural
  Networks}. In \bibinfo{booktitle}{\emph{Conference on Neural Information
  Processing Systems}}.
\newblock


\bibitem[Goodfellow et~al\mbox{.}(2014)]%
        {goodfellow2014generative}
\bibfield{author}{\bibinfo{person}{Ian Goodfellow}, \bibinfo{person}{Jean
  Pouget-Abadie}, \bibinfo{person}{Mehdi Mirza}, \bibinfo{person}{Bing Xu},
  \bibinfo{person}{David Warde-Farley}, \bibinfo{person}{Sherjil Ozair},
  \bibinfo{person}{Aaron Courville}, {and} \bibinfo{person}{Yoshua Bengio}.}
  \bibinfo{year}{2014}\natexlab{}.
\newblock \showarticletitle{Generative adversarial nets}. In
  \bibinfo{booktitle}{\emph{Conference on Neural Information Processing
  Systems}}.
\newblock


\bibitem[Hartley and Zisserman(2003)]%
        {hartley2003multiple}
\bibfield{author}{\bibinfo{person}{Richard Hartley} {and}
  \bibinfo{person}{Andrew Zisserman}.} \bibinfo{year}{2003}\natexlab{}.
\newblock \bibinfo{booktitle}{\emph{Multiple View Geometry in Computer
  Vision}}.
\newblock \bibinfo{publisher}{Cambridge University Press}.
\newblock


\bibitem[Heusel et~al\mbox{.}(2017)]%
        {heusel17}
\bibfield{author}{\bibinfo{person}{Martin Heusel}, \bibinfo{person}{Hubert
  Ramsauer}, \bibinfo{person}{Thomas Unterthiner}, \bibinfo{person}{Bernhard
  Nessler}, {and} \bibinfo{person}{Sepp Hochreiter}.}
  \bibinfo{year}{2017}\natexlab{}.
\newblock \showarticletitle{GANs Trained by a Two Time-Scale Update Rule
  Converge to a Local Nash Equilibrium}. In
  \bibinfo{booktitle}{\emph{Conference on Neural Information Processing
  Systems}}.
\newblock


\bibitem[Ho et~al\mbox{.}(2020)]%
        {ddpm}
\bibfield{author}{\bibinfo{person}{Jonathan Ho}, \bibinfo{person}{Ajay Jain},
  {and} \bibinfo{person}{Pieter Abbeel}.} \bibinfo{year}{2020}\natexlab{}.
\newblock \showarticletitle{Denoising Diffusion Probabilistic Models}. In
  \bibinfo{booktitle}{\emph{Advances in Neural Information Processing
  Systems}}.
\newblock


\bibitem[Ho and Salimans(2021)]%
        {ho2021classifierfree}
\bibfield{author}{\bibinfo{person}{Jonathan Ho} {and} \bibinfo{person}{Tim
  Salimans}.} \bibinfo{year}{2021}\natexlab{}.
\newblock \showarticletitle{Classifier-Free Diffusion Guidance}. In
  \bibinfo{booktitle}{\emph{NeurIPS Workshop on Deep Generative Models and
  Downstream Applications}}.
\newblock


\bibitem[Isola et~al\mbox{.}(2017)]%
        {isola2017image}
\bibfield{author}{\bibinfo{person}{Phillip Isola}, \bibinfo{person}{Jun-Yan
  Zhu}, \bibinfo{person}{Tinghui Zhou}, {and} \bibinfo{person}{Alexei~A.
  Efros}.} \bibinfo{year}{2017}\natexlab{}.
\newblock \showarticletitle{Image-to-Image Translation with Conditional
  Adversarial Nets}. In \bibinfo{booktitle}{\emph{IEEE Conference on Computer
  Vision and Pattern Recognition}}.
\newblock


\bibitem[Isoyama et~al\mbox{.}(2021)]%
        {isoyama2021effects}
\bibfield{author}{\bibinfo{person}{Naoya Isoyama}, \bibinfo{person}{Yamato
  Sakuragi}, \bibinfo{person}{Tsutomu Terada}, {and} \bibinfo{person}{Masahiko
  Tsukamoto}.} \bibinfo{year}{2021}\natexlab{}.
\newblock \showarticletitle{Effects of Augmented Reality Object and Texture
  Presentation on Walking Behavior}.
\newblock \bibinfo{journal}{\emph{Electronics}} \bibinfo{volume}{10},
  \bibinfo{number}{6} (\bibinfo{year}{2021}).
\newblock


\bibitem[Jetchev et~al\mbox{.}(2016)]%
        {SGAN16}
\bibfield{author}{\bibinfo{person}{Nikolay Jetchev}, \bibinfo{person}{Urs~M.
  Bergmann}, {and} \bibinfo{person}{Roland Vollgraf}.}
  \bibinfo{year}{2016}\natexlab{}.
\newblock \showarticletitle{Texture Synthesis with Spatial Generative
  Adversarial Networks}.
\newblock \bibinfo{journal}{\emph{CoRR}}  \bibinfo{volume}{abs/1611.08207}
  (\bibinfo{year}{2016}).
\newblock


\bibitem[Johnson et~al\mbox{.}(2016)]%
        {jognson_perceptual_2016}
\bibfield{author}{\bibinfo{person}{Justin Johnson}, \bibinfo{person}{Alexandre
  Alahi}, {and} \bibinfo{person}{Li Fei-Fei}.} \bibinfo{year}{2016}\natexlab{}.
\newblock \showarticletitle{Perceptual Losses for Real-Time Style Transfer and
  Super-Resolution}. In \bibinfo{booktitle}{\emph{European Conference on
  Computer Vision}}.
\newblock


\bibitem[Kingma and Ba(2015)]%
        {kingma2015adam}
\bibfield{author}{\bibinfo{person}{Diederik~P. Kingma} {and}
  \bibinfo{person}{Jimmy Ba}.} \bibinfo{year}{2015}\natexlab{}.
\newblock \showarticletitle{Adam: A Method for Stochastic Optimization}. In
  \bibinfo{booktitle}{\emph{International Conference on Learning
  Representations}}.
\newblock


\bibitem[Kwitt and Meerwald(2008)]%
        {salzburg_texture_db}
\bibfield{author}{\bibinfo{person}{Roland Kwitt} {and} \bibinfo{person}{Peter
  Meerwald}.} \bibinfo{year}{2008}\natexlab{}.
\newblock \bibinfo{title}{Salzburg Texture Image Database}.
\newblock \bibinfo{howpublished}{Online}.
\newblock


\bibitem[Li et~al\mbox{.}(2022a)]%
        {li2022mat}
\bibfield{author}{\bibinfo{person}{Wenbo Li}, \bibinfo{person}{Zhe Lin},
  \bibinfo{person}{Kun Zhou}, \bibinfo{person}{Lu Qi}, \bibinfo{person}{Yi
  Wang}, {and} \bibinfo{person}{Jiaya Jia}.} \bibinfo{year}{2022}\natexlab{a}.
\newblock \showarticletitle{MAT: Mask-Aware Transformer for Large Hole Image
  Inpainting}. In \bibinfo{booktitle}{\emph{IEEE Conference on Computer Vision
  and Pattern Recognition}}.
\newblock


\bibitem[Li et~al\mbox{.}(2022b)]%
        {li2022scraping}
\bibfield{author}{\bibinfo{person}{Xueting Li}, \bibinfo{person}{Xiaolong
  Wang}, \bibinfo{person}{Ming-Hsuan Yang}, \bibinfo{person}{Alexei~A. Efros},
  {and} \bibinfo{person}{Sifei Liu}.} \bibinfo{year}{2022}\natexlab{b}.
\newblock \showarticletitle{Scraping Textures from Natural Images for Synthesis
  and Editing}. In \bibinfo{booktitle}{\emph{European Conference on Computer
  Vision}}.
\newblock


\bibitem[Li et~al\mbox{.}(2017)]%
        {li2017diversified}
\bibfield{author}{\bibinfo{person}{Yijun Li}, \bibinfo{person}{Chen Fang},
  \bibinfo{person}{Jimei Yang}, \bibinfo{person}{Zhaowen Wang},
  \bibinfo{person}{Xin Lu}, {and} \bibinfo{person}{Ming-Hsuan Yang}.}
  \bibinfo{year}{2017}\natexlab{}.
\newblock \showarticletitle{Diversified Texture Synthesis with Feed-forward
  Networks}. In \bibinfo{booktitle}{\emph{IEEE Conference on Computer Vision
  and Pattern Recognition}}.
\newblock


\bibitem[Liu et~al\mbox{.}(2018)]%
        {liu2018partial}
\bibfield{author}{\bibinfo{person}{Guilin Liu}, \bibinfo{person}{Fitsum~A.
  Reda}, \bibinfo{person}{Kevin~J. Shih}, \bibinfo{person}{Ting-Chun Wang},
  \bibinfo{person}{Andrew Tao}, {and} \bibinfo{person}{Bryzan Catanzaro}.}
  \bibinfo{year}{2018}\natexlab{}.
\newblock \showarticletitle{Image Inpainting for Irregular Holes Using Partial
  Convolutions}. In \bibinfo{booktitle}{\emph{European Conference on Computer
  Vision}}.
\newblock


\bibitem[Liu et~al\mbox{.}(2020)]%
        {transposer}
\bibfield{author}{\bibinfo{person}{Guilin Liu}, \bibinfo{person}{Rohan Taori},
  \bibinfo{person}{Ting-Chun Wang}, \bibinfo{person}{Zhiding Yu},
  \bibinfo{person}{Shiqiu Liu}, \bibinfo{person}{Fitsum~A. Reda},
  \bibinfo{person}{Karan Sapra}, \bibinfo{person}{Andrew Tao}, {and}
  \bibinfo{person}{Bryan Catanzaro}.} \bibinfo{year}{2020}\natexlab{}.
\newblock \showarticletitle{Transposer: Universal Texture Synthesis Using
  Feature Maps as Transposed Convolution Filter}.
\newblock \bibinfo{journal}{\emph{CoRR}}  \bibinfo{volume}{abs/2007.07243}
  (\bibinfo{year}{2020}).
\newblock


\bibitem[Mallikarjuna et~al\mbox{.}(2006)]%
        {Mallikarjuna2006THEK2}
\bibfield{author}{\bibinfo{person}{P.~B. Mallikarjuna},
  \bibinfo{person}{Alireza~Tavakoli Targhi}, \bibinfo{person}{Mario Fritz},
  \bibinfo{person}{Eric Hayman}, \bibinfo{person}{Barbara Caputo}, {and}
  \bibinfo{person}{J.~O. Eklundh}.} \bibinfo{year}{2006}\natexlab{}.
\newblock \showarticletitle{THE KTH-TIPS 2 database}.
\newblock


\bibitem[Mardani et~al\mbox{.}(2020)]%
        {mardani2020neural}
\bibfield{author}{\bibinfo{person}{Morteza Mardani}, \bibinfo{person}{Guilin
  Liu}, \bibinfo{person}{Aysegul Dundar}, \bibinfo{person}{Shiqiu Liu},
  \bibinfo{person}{Andrew Tao}, {and} \bibinfo{person}{Bryan Catanzaro}.}
  \bibinfo{year}{2020}\natexlab{}.
\newblock \showarticletitle{Neural FFTs for Universal Texture Image Synthesis}.
  In \bibinfo{booktitle}{\emph{Advances in Neural Information Processing
  Systems}}.
\newblock


\bibitem[Mehdi~Mirza(2014)]%
        {cGAN}
\bibfield{author}{\bibinfo{person}{Simon~Osindero Mehdi~Mirza}.}
  \bibinfo{year}{2014}\natexlab{}.
\newblock \showarticletitle{Conditional Generative Adversarial Nets}.
\newblock \bibinfo{journal}{\emph{CoRR}}  \bibinfo{volume}{abs/1411.1784}
  (\bibinfo{year}{2014}).
\newblock


\bibitem[Picard et~al\mbox{.}(2010)]%
        {vistex_texture}
\bibfield{author}{\bibinfo{person}{Rosalind Picard}, \bibinfo{person}{Chris
  Graczyk}, \bibinfo{person}{Steve Mann}, \bibinfo{person}{Josh Wachman},
  \bibinfo{person}{Len Picard}, {and} \bibinfo{person}{Lee Campbell}.}
  \bibinfo{year}{2010}\natexlab{}.
\newblock \bibinfo{title}{Vistex Vision Texture Database}.
\newblock \bibinfo{howpublished}{Online}.
\newblock


\bibitem[Riba et~al\mbox{.}(2020)]%
        {eriba2019kornia}
\bibfield{author}{\bibinfo{person}{E. Riba}, \bibinfo{person}{D. Mishkin},
  \bibinfo{person}{D. Ponsa}, \bibinfo{person}{E. Rublee}, {and}
  \bibinfo{person}{G. Bradski}.} \bibinfo{year}{2020}\natexlab{}.
\newblock \showarticletitle{Kornia: an Open Source Differentiable Computer
  Vision Library for PyTorch}. In \bibinfo{booktitle}{\emph{WACV}}.
\newblock


\bibitem[Rombach et~al\mbox{.}(2022)]%
        {rombach2022ldm}
\bibfield{author}{\bibinfo{person}{Robin Rombach}, \bibinfo{person}{Andreas
  Blattmann}, \bibinfo{person}{Dominik Lorenz}, \bibinfo{person}{Patrick
  Esser}, {and} \bibinfo{person}{Bjorn Ommer}.}
  \bibinfo{year}{2022}\natexlab{}.
\newblock \showarticletitle{High-Resolution Image Synthesis with Latent
  Diffusion Models}. In \bibinfo{booktitle}{\emph{IEEE Conference on Computer
  Vision and Pattern Recognition}}.
\newblock


\bibitem[Saharia et~al\mbox{.}(2022)]%
        {saharia2022palette}
\bibfield{author}{\bibinfo{person}{Chitwan Saharia}, \bibinfo{person}{William
  Chan}, \bibinfo{person}{Huiwen Chang}, \bibinfo{person}{Chris~A. Lee},
  \bibinfo{person}{Jonathan Ho}, \bibinfo{person}{Tim Salimans},
  \bibinfo{person}{David~J. Fleet}, {and} \bibinfo{person}{Mohammad Norouzi}.}
  \bibinfo{year}{2022}\natexlab{}.
\newblock \showarticletitle{Palette: Image-to-Image Diffusion Models}. In
  \bibinfo{booktitle}{\emph{SIGGRAPH '22: ACM SIGGRAPH Conference
  Proceedings}}.
\newblock


\bibitem[Saharia et~al\mbox{.}(2023)]%
        {sr3}
\bibfield{author}{\bibinfo{person}{Chitwan Saharia}, \bibinfo{person}{Jonathan
  Ho}, \bibinfo{person}{William Chan}, \bibinfo{person}{Tim Salimans},
  \bibinfo{person}{David~J. Fleet}, {and} \bibinfo{person}{Mohammad Norouzi}.}
  \bibinfo{year}{2023}\natexlab{}.
\newblock \showarticletitle{Image Super-Resolution via Iterative Refinement}.
\newblock \bibinfo{journal}{\emph{IEEE Transactions on Pattern Analysis and
  Machine Intelligence}} \bibinfo{volume}{45}, \bibinfo{number}{4}
  (\bibinfo{year}{2023}).
\newblock


\bibitem[Sharan et~al\mbox{.}(2014)]%
        {sharan2014accuracy}
\bibfield{author}{\bibinfo{person}{Lavanya Sharan}, \bibinfo{person}{Ce Liu},
  \bibinfo{person}{Ruth Rosenholtz}, {and} \bibinfo{person}{Edward~H.
  Adelson}.} \bibinfo{year}{2014}\natexlab{}.
\newblock \showarticletitle{Accuracy and Speed of Material Categorization in
  Real-World Images}.
\newblock \bibinfo{journal}{\emph{Journal of Vision}} \bibinfo{volume}{14},
  \bibinfo{number}{9} (\bibinfo{year}{2014}), \bibinfo{pages}{article 12}.
\newblock


\bibitem[Sohl-Dickstein et~al\mbox{.}(2015)]%
        {sohl2015deep}
\bibfield{author}{\bibinfo{person}{Jascha Sohl-Dickstein},
  \bibinfo{person}{Eric Weiss}, \bibinfo{person}{Niru Maheswaranathan}, {and}
  \bibinfo{person}{Surya Ganguli}.} \bibinfo{year}{2015}\natexlab{}.
\newblock \showarticletitle{Deep Unsupervised Learning Using Nonequilibrium
  Thermodynamics}. In \bibinfo{booktitle}{\emph{International Conference on
  Machine Learning}}.
\newblock


\bibitem[Song et~al\mbox{.}(2021a)]%
        {song2021denoising}
\bibfield{author}{\bibinfo{person}{Jiaming Song}, \bibinfo{person}{Chenlin
  Meng}, {and} \bibinfo{person}{Stefano Ermon}.}
  \bibinfo{year}{2021}\natexlab{a}.
\newblock \showarticletitle{Denoising Diffusion Implicit Models}. In
  \bibinfo{booktitle}{\emph{International Conference on Learning
  Representations}}.
\newblock


\bibitem[Song et~al\mbox{.}(2021b)]%
        {song2021score}
\bibfield{author}{\bibinfo{person}{Yang Song}, \bibinfo{person}{Jascha
  Sohl-Dickstein}, \bibinfo{person}{Diederik~P. Kingma},
  \bibinfo{person}{Abhishek Kumar}, \bibinfo{person}{Stefano Ermon}, {and}
  \bibinfo{person}{Ben Poole}.} \bibinfo{year}{2021}\natexlab{b}.
\newblock \showarticletitle{Score-Based Generative Modeling through Stochastic
  Differential Equations}. In \bibinfo{booktitle}{\emph{International
  Conference on Learning Representations}}.
\newblock


\bibitem[van~den Oord et~al\mbox{.}(2017)]%
        {vandenOord2017neural}
\bibfield{author}{\bibinfo{person}{Aaron van~den Oord}, \bibinfo{person}{Oriol
  Vinyals}, {and} \bibinfo{person}{Koray Kavukcuoglu}.}
  \bibinfo{year}{2017}\natexlab{}.
\newblock \showarticletitle{Neural Discrete Representation Learning}. In
  \bibinfo{booktitle}{\emph{Conference on Neural Information Processing
  Systems}}.
\newblock


\bibitem[Verbin and Zickler(2020)]%
        {verbin2020toward}
\bibfield{author}{\bibinfo{person}{Dor Verbin} {and} \bibinfo{person}{Todd
  Zickler}.} \bibinfo{year}{2020}\natexlab{}.
\newblock \showarticletitle{Toward a Universal Model for Shape from Texture}.
  In \bibinfo{booktitle}{\emph{IEEE Conference on Computer Vision and Pattern
  Recognition}}.
\newblock


\bibitem[Wang et~al\mbox{.}(2004)]%
        {wang2004image}
\bibfield{author}{\bibinfo{person}{Zhou Wang}, \bibinfo{person}{Alan~C. Bovik},
  \bibinfo{person}{Hamid~R. Sheikh}, {and} \bibinfo{person}{Eero~P.
  Simoncelli}.} \bibinfo{year}{2004}\natexlab{}.
\newblock \showarticletitle{Image Quality Assessment: From Error Visibility to
  Structural Similarity}.
\newblock \bibinfo{journal}{\emph{IEEE Transactions on Image Processing}}
  \bibinfo{volume}{13}, \bibinfo{number}{4} (\bibinfo{year}{2004}),
  \bibinfo{pages}{600--612}.
\newblock


\bibitem[Wei et~al\mbox{.}(2009)]%
        {wei2009state}
\bibfield{author}{\bibinfo{person}{Li-yi Wei}, \bibinfo{person}{Sylvain
  Lefebvre}, \bibinfo{person}{Vivek Kwatra}, {and} \bibinfo{person}{Greg
  Turk}.} \bibinfo{year}{2009}\natexlab{}.
\newblock \showarticletitle{State of the Art in Example-based Texture
  Synthesis}. In \bibinfo{booktitle}{\emph{Eurographics 2009 - State of the Art
  Reports}}.
\newblock


\bibitem[Wei and Levoy(2000)]%
        {wei2000fast}
\bibfield{author}{\bibinfo{person}{Li-Yi Wei} {and} \bibinfo{person}{Marc
  Levoy}.} \bibinfo{year}{2000}\natexlab{}.
\newblock \showarticletitle{Fast Texture Synthesis Using Tree-Structured Vector
  Quantization}. In \bibinfo{booktitle}{\emph{SIGGRAPH '00: Proceedings of the
  27th Annual Conference on Computer Graphics and Interactive Techniques}}.
  \bibinfo{pages}{479--488}.
\newblock


\bibitem[Wu et~al\mbox{.}(2018)]%
        {wu2018automatic}
\bibfield{author}{\bibinfo{person}{Huisi Wu}, \bibinfo{person}{Xiaomeng Lyu},
  {and} \bibinfo{person}{Zhenkun Wen}.} \bibinfo{year}{2018}\natexlab{}.
\newblock \showarticletitle{Automatic texture exemplar extraction based on
  global and local textureness measures}.
\newblock \bibinfo{journal}{\emph{Computational Visual Media}}
  \bibinfo{volume}{4} (\bibinfo{year}{2018}), \bibinfo{pages}{173--184}.
\newblock


\bibitem[Wu et~al\mbox{.}(2021)]%
        {wu2021deep}
\bibfield{author}{\bibinfo{person}{Huisi Wu}, \bibinfo{person}{Wei Yan},
  \bibinfo{person}{Ping Li}, {and} \bibinfo{person}{Zhenkun Wen}.}
  \bibinfo{year}{2021}\natexlab{}.
\newblock \showarticletitle{Deep Texture Exemplar Extraction Based on Trimmed
  T-CNN}.
\newblock \bibinfo{journal}{\emph{IEEE Transactions on Multimedia}}
  \bibinfo{volume}{23} (\bibinfo{year}{2021}), \bibinfo{pages}{4502--4514}.
\newblock


\bibitem[Yu et~al\mbox{.}(2019)]%
        {yu2019freeform}
\bibfield{author}{\bibinfo{person}{Jiahui Yu}, \bibinfo{person}{Zhe Lin},
  \bibinfo{person}{Jimei Yang}, \bibinfo{person}{Xiaohui Shen},
  \bibinfo{person}{Xin Lu}, {and} \bibinfo{person}{Thomas Huang}.}
  \bibinfo{year}{2019}\natexlab{}.
\newblock \showarticletitle{Free-Form Image Inpainting with Gated Convolution}.
  In \bibinfo{booktitle}{\emph{International Conference on Computer Vision}}.
\newblock


\bibitem[Zhang et~al\mbox{.}(2019)]%
        {zhang2019selfattention}
\bibfield{author}{\bibinfo{person}{Han Zhang}, \bibinfo{person}{Ian
  Goodfellow}, \bibinfo{person}{Dimitris Metaxas}, {and}
  \bibinfo{person}{Augustus Odena}.} \bibinfo{year}{2019}\natexlab{}.
\newblock \showarticletitle{Self-Attention Generative Adversarial Networks}. In
  \bibinfo{booktitle}{\emph{International Conference on Machine Learning}}.
\newblock


\bibitem[Zhang et~al\mbox{.}(2018)]%
        {zhang2018unreasonable}
\bibfield{author}{\bibinfo{person}{Richard Zhang}, \bibinfo{person}{Phillip
  Isola}, \bibinfo{person}{Alexei~A. Efros}, \bibinfo{person}{Eli Shechtman},
  {and} \bibinfo{person}{Oliver Wang}.} \bibinfo{year}{2018}\natexlab{}.
\newblock \showarticletitle{The Unreasonable Effectiveness of Deep Features as
  a Perceptual Metric}. In \bibinfo{booktitle}{\emph{IEEE Conference on
  Computer Vision and Pattern Recognition}}.
\newblock


\bibitem[Zhou et~al\mbox{.}(2018)]%
        {nonstat}
\bibfield{author}{\bibinfo{person}{Yang Zhou}, \bibinfo{person}{Zhen Zhu},
  \bibinfo{person}{Xiang Bai}, \bibinfo{person}{Dani Lischinski},
  \bibinfo{person}{Daniel Cohen-Or}, {and} \bibinfo{person}{Hui Huang}.}
  \bibinfo{year}{2018}\natexlab{}.
\newblock \showarticletitle{Non-Stationary Texture Synthesis by Adversarial
  Expansion}.
\newblock \bibinfo{journal}{\emph{ACM Transactions on Graphics (Proceedings of
  SIGGRAPH)}} \bibinfo{volume}{37}, \bibinfo{number}{4} (\bibinfo{year}{2018}),
  \bibinfo{numpages}{13}~pages.
\newblock


\bibitem[Zhu et~al\mbox{.}(2017)]%
        {zhu2017unpaired}
\bibfield{author}{\bibinfo{person}{Jun-Yan Zhu}, \bibinfo{person}{Taesung
  Park}, \bibinfo{person}{Phillip Isola}, {and} \bibinfo{person}{Alexei~A.
  Efros}.} \bibinfo{year}{2017}\natexlab{}.
\newblock \showarticletitle{Unpaired Image-to-Image Translation using
  Cycle-Consistent Adversarial Networks}. In
  \bibinfo{booktitle}{\emph{International Conference on Computer Vision}}.
\newblock


\end{thebibliography}
